# Revisiting thermal conductivity and interface conductance at the nanoscale

B. Davier[1,2], P. Dollfus[1], S. Volz[3], J. Shiomi[2], J. Saint-Martin[1]

[1] Université Paris-Saclay, CNRS, Centre de Nanosciences et de Nanotechnologies, 91120, Palaiseau, France

[2] Department of Mechanical Engineering, The University of Tokyo, Tokyo 113-856, Japan

[3] LIMMS UMI 2820, The University of Tokyo-CNRS, 4-6-1 Komaba, Meguro-ku, Tokyo 153-8505, Japan

## Abstract

A semi-analytical model for studying thermal transport at the nanoscale, able to accurately describe both the effect of out of equilibrium transport and the thermal transfer at interfaces, is presented. Our approach is based on the definition of pseudo local temperatures distinguishing the phonon populations according to the direction of their velocity. This formalism leads to a complete set of equations capturing the heat transfer in nanostructures even in the case of hetero-structures. This model only requires introducing a new intrinsic thermal parameter called ballistic thermal conductance and a geometric one called the effective thermal conductivity. Finally, this model is able to reproduce accurately advanced numerical results of Monte Carlo simulation for phonons in all phonon transport regime: diffusive (as the Fourier heat transport regime is included), ballistic, and intermediate ones even if thermal interface are involved. This formalism should provide new insights in the interpretation of experimental measurements.

## I  Introduction

The standard formalism of heat transport in solids built in the nineteen century by Fourier [1], is based on two material properties: the heat capacity and the thermal conductivity. Using quantum mechanics and kinetic theory for phonons, the quantized heat carriers, Peierls established in 1929 [2] the heat theory that is still in used nowadays. By using this complex theoretical framework, the heat transport parameters can be computed numerically via ab-initio approaches generally based on Density Functional Theory (DFT) [3], [4] and are now in a remarkable agreement with experimental measurements [5]. In bulk materials, spatial dimensions are, by definition, much longer than the mean free path of phonons, and the diffusive heat transport regime that takes place is perfectly captured by Fourier's formalism. In such systems, phonon energy distributions remain close to their equilibrium distribution, the Bose-Einstein statistics.

These criteria are not valid in many recent applications based on nanosystems [6]. For instance, current transistors are nanometer-long [7], and materials of interest in thermoelectrics are nanostructured, such as superlattices [8] or stacks of 2D materials [9], [10]. In these systems the dimension of which are of the same order of magnitude as or even smaller than the phonon mean free path, the phonon transport is out of equilibrium as the number of scattering events encountered by phonons is not sufficient to let them recover their equilibrium distribution. As the Fourier heat formalism reaches its limit of validity at the nanoscale, advanced numerical methods are mandatory to investigate heat transfer. The most common approaches are based on the Boltzmann transport formalism for phonons such as the Monte Carlo simulation [11]–[14], others are based on the Non Equilibrium Green's Functions framework [15]–[17], while some others calculate the trajectory of atoms by using Molecular Dynamics [18], [19]. However, a simple analytical modeling of heat transfer based on a set of a few parameters relevant at the nanometer scale is highly desirable.

Besides, much attention has been given to the investigation of heat transport across interfaces since the pioneering works of Kapitza [20]. In order to study the specific case of solid-solid interfaces, Little [21] adapted the Acoustic Mismatch Model (AMM) and later Swartz [22] developed the Diffusive Mismatch Model (DMM). Even if their underlying assumptions seem very different, these two models are based on a Landauer approach considering the phonons emitted by ideal thermostats. However, though effective for many problems, they both lead to the so called virtual interface paradox as they predict a non-zero thermal interface resistance (or a finite conductance) in the case of an imaginary fully transparent interface (with a transmission of 1) located inside any homogenous material [22].

To overcome this paradox the contribution of out of equilibrium phonons have been included by Simons [23], Chen [24] or Merabia et al. [25] for instance. By using a solution of the Boltzmann transport equation in the linear regime (close to equilibrium), they derived a corrective term to modify the usual pseudo temperatures at the interface. However, these temperatures are generally phonon mode dependent [26], leading to models difficult to handle and even unable to correctly capture the fully ballistic transport regime.

The present work aims at introducing a simple model based on a definitions of an effective thermal conductivity and an interface thermal conductance which generalize the common macroscopic parameters of the Fourier formalism and extend their validity at the nanoscale in homogeneous and inhomogeneous systems. These parameters are intrinsically related to two effective (not phonon mode dependent) pseudo temperatures distinguishing the populations of phonons according to the (positive or negative) direction of their velocity. The proposed analytical model is benchmarked with advanced Monte Carlo simulation for phonons [14].

## II   Homogenous system

### II.1.   Thermal conductivity

Within the framework of Fourier's law, the thermal conductivity in diffusive regime (i.e. in a system longer than the mean free path) is the proportionality factor between the heat flux density $Q$ and the temperature gradient along the transport direction $x$:

$$Q = -\kappa_{\text{diffusive}} \frac{\partial T}{\partial x} \qquad (1)$$

According to Peierls model the bulk thermal conductivity of a material is expressed as [25]

$$\kappa = \frac{\Omega}{(2\pi)^3} \sum_s \hbar \omega_s |v_{s,x}| l_s \frac{\partial f_{\text{BE}}}{\partial T}(\omega_s, \bar{T}) \qquad (2)$$

where $\bar{T}$ is the average temperature in the system, and $l_s$ is the mean free path parameter for a phonon in state $s$. The quantities $\omega_s$ and $v_s$ are the angular frequency and velocity of state $s$. $\Omega$ is the volume of the considered reciprocal space and $f_{\text{BE}}$ is the Bose-Einstein distribution.

In the diffusive regime, a local model of $l_s$ based only of the properties of the material is sufficient:

$$l_{s,\text{diffusive}} = \frac{|v_{s,x}|}{\lambda_s}$$

where $\lambda_s$ is the rate of phonon-phonon scattering for a phonon in state $s$ within a relaxation time approximation. The diffusive (i.e. bulk) conductivity is thus:

$$\kappa_{\text{diffusive}} = \frac{\Omega}{(2\pi)^3} \sum_s \hbar \omega_s \frac{|v_{s,x}|^2}{\lambda_s} \frac{\partial f_{\text{BE}}}{\partial T}(\omega_s, \bar{T}) \qquad (3)$$

However, in non-diffusive regimes, this relation is no longer representative of the heat transport. This is particularly true in ballistic systems in which the temperature gradient vanishes.

### II.2.   Ballistic conductance

To investigate a ballistic system, i.e. in which no scattering mechanism occurs, the standard framework is the Landauer formalism. The total heat flux exchanged between two thermostats (a cold one and a hot one) through a perfect channel (having a transmission of one) can be written as:

$$Q = Q_{\text{hot\_contact}} - Q_{\text{cold\_contact}} \qquad (4)$$
$$= \frac{1}{2} \frac{\Omega}{(2\pi)^3} \sum_s \hbar\omega_s |v_{s,x}| \left( f_{BE}(\omega_s, T_{\text{hot\_contact}}) - f_{BE}(\omega_s, T_{\text{cold\_contact}}) \right)$$

For a small temperature difference $\Delta T_{\text{contacts}} = T_{\text{hot\_contact}} - T_{\text{cold\_contact}}$, the first order gradient expansion gives:

$$f_{BE}(\omega_s, T_{\text{hot\_contact}}) - f_{BE}(\omega_s, T_{\text{cold\_contact}}) \approx \frac{\partial f_{BE}}{\partial T} \Delta T_{\text{contacts}}$$

Then, the heat flux density (8) may thus be rewritten by defining a ballistic thermal conductance $G_{\text{ballistic}}$ as follows:

$$Q = \left( \frac{1}{2} \cdot \frac{\Omega}{(2\pi)^3} \sum_s \hbar\omega_s |v_{s,x}| \times \frac{\partial f_{BE}}{\partial T} \right) \Delta T_{\text{contacts}} = G_{\text{ballistic}} \times \Delta T_{\text{contacts}} \qquad (5)$$

It is worth noting that in a ballistic system [27], [28], the conductance is not size ($L$) dependent but depends on material phonon dispersion.

## II.3. Effective thermal conductivity

To bridge the gap between diffusive and ballistic heat transport regimes, we propose to introduce an effective thermal conductivity $\kappa_{\text{effective}}$ that is not defined from the temperature gradient but by using the temperature difference between the incoming phonons on each sides considering the whole system, i.e. the temperature $\Delta T_{\text{contacts}}$ as defined above. In a system of length $L$ between the thermostats, we define $\kappa_{\text{effective}}$ as follows:

$$Q = \kappa_{\text{effective}} \frac{\Delta T_{\text{contacts}}}{L} \qquad (6)$$

By using this definition, $\kappa_{\text{effective}}$ and $\kappa_{\text{diffusive}}$ are equivalent in the diffusive regime, i.e. in systems much longer than the mean free path of phonons.

Besides, in a ballistic system where the temperature gradient is zero, a ballistic thermal conductivity $\kappa_{\text{ballistic}}$ cannot be defined by using the usual definition of Eq. 1. However, combining Eq. 5 and 6 gives:

$$\kappa_{\text{ballistic}} = \frac{\Omega}{(2\pi)^3} \sum_s \hbar\omega_s |v_{s,x}| \cdot \frac{L}{2} \cdot \frac{\partial f_{BE}}{\partial T}(\omega_s, \bar{T}) = L\, G_{\text{ballistic}} \qquad (7)$$

Interestingly, by using Eq. 2 to identify the ballistic mean free path in Eq. 7, it yields: $l_{s,\text{ballistic}} = \frac{L}{2}$. Indeed, in ballistic regime, the mean free path only depends on the system dimension. The ballistic thermal conductivity is thus proportional to the distance between the thermostats $L$.

The simplest expression of the effective thermal conductivity in the intermediate transport regime (i.e. between diffusive and ballistic ones) depending on both the properties of the material and the structure is a non-spectral Matthiessen rule as follows:

$$\frac{1}{\kappa_{\text{effective,Matthiessen}}} = \frac{1}{\kappa_{\text{ballistic}}} + \frac{1}{\kappa_{\text{diffusive}}} \qquad (8)$$

To provide a more accurate estimation of the effective conductivity, a spectral Matthiessen summation for the mean free path must be applied by considering Eq. 3. Additionally, the the spectral mean free path should depend on the geometry. For instance, the effective thermal conductivity of a homogeneous system of length $L$ considering nanofilms in cross-plane (CPNF) configuration is (cf. [14]):

$$\kappa_{\text{effective,CPNF}} = \frac{\Omega}{(2\pi)^3} \sum_s \hbar\omega_s |v_{s,x}| l_{s,\text{CPNF}} \frac{\partial f_{BE}}{\partial T}(\omega_s, \bar{T}) \text{ with } l_{s,\text{CPNF}} = \frac{|v_{s,x}|}{\lambda_s + 2\frac{|v_{s,x}|}{L}} \qquad (9)$$

The thermal conductivities obtained by these different models are plotted in Figure 1 for Silicon nanofilms, as a function of their length $L$, and compared with the effective thermal conductivity extracted from Monte Carlo simulations using (6). As expected, all models converge asymptotically to the ballistic and diffusive limits. However, the simple non-spectral Matthiessen approximation (Eq. 8) differs from the MC results in the intermediate regime, up to 60% at $L = 200$nm. The spectral model (Eq. 9, blue solid line) agrees with the MC results in the full range of $L$.

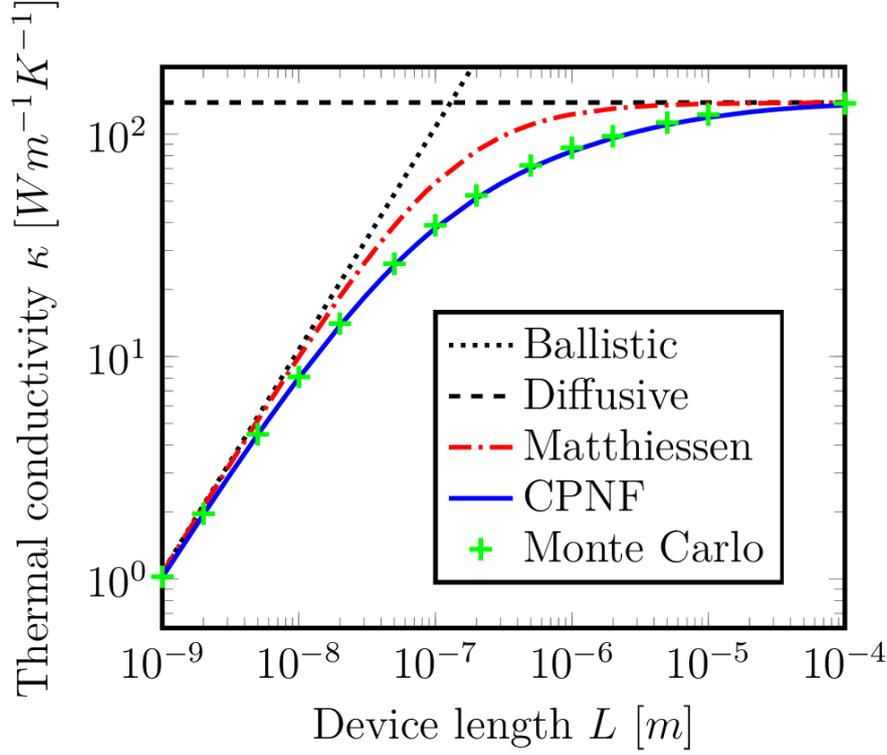

*Figure 1: Thermal conductivities as a function of the device length for for different models. Green crosses: Monte Carlo results, black dotted line: Ballitic model (Eq. 7), black dashed line: Diffusive model (Eq. 3), red dashed line: Matthiessen model (Eq. 8), blue line: effective conductivity (Eq. 9).*

## II.4. Hemispherical temperatures

The concept of temperature, strongly linked to the equilibrium Bose-Einstein statistics of phonons, is ill-defined in non-diffusive heat transport regime because the distribution of phonons may be far from equilibrium. It is however always convenient to extract a temperature as this parameter is widely used in the previous definitions see for instance Eq. 1. Usually, a pseudo temperature $T$ is defined as the temperature leading to the same phonon energy density as the actual density[14] but using an equilibrium distribution of phonons at this temperature $T$.

However, to use the definition of the effective conductivity in Eq. 6 or to treat the case of heat transport through interfaces, it is also useful to separate the phonon distribution in two parts, i.e. one part related to the phonons having a positive velocity component along the main heat flux and another part related to the remaining phonons having a negative velocity. Consequently, the "hemispherical" temperatures $T^+$ and $T^-$ are defined as follows:

$$E^{+/-}(T^{+/-}) = \frac{\Omega}{(2\pi)^3} \sum_{\substack{s \\ v_s>0 \, / \, v_s<0}} \hbar\omega_s f_{\text{BE}}(\omega_s, T^{+/-}) \tag{10}$$

where $E^{+/-}$ is the energy density of actual positive-going/negative-going phonon distribution. The summation on the reciprocal states of Eq. (10) is made over all states $s$ of the Brillouin zone.

In the following discussion for any homogeneous section of a linear system, two different temperature differences are used. The first one, called $\Delta T_{\text{contacts}}$, is the difference of temperature between incoming phonons onto this section, i.e. between right-going phonons at the left boundary and left-going phonons at the right boundary. For a section of length $L$, we thus have $\Delta T_{\text{contacts}} = T^+(x=0) - T^-(x=L)$. The second one is a local temperature difference $\Delta T_{\text{local}}$ defined at any position $x$ as the difference $\Delta T_{\text{local}}(x) = T^+(x) - T^-(x)$.

## II.5. Heat flux and fundamental relationships

By introducing the local hemispherical temperatures $T^+$ and $T^-$ in a homogeneous section of a system, the heat flux density may also be written naturally in the framework of a Landauer formalism at any position $x$:

$$Q = Q|_{v_x>0} - Q|_{v_x<0} = \frac{1}{2}\frac{\Omega}{(2\pi)^3}\sum_s \hbar\omega_s |v_{s,x}| \left( f_{BE}(\omega_s, T^+(x)) - f_{BE}(\omega_s, T^-(x)) \right) \tag{11}$$

For a small temperature difference $\Delta T_{\text{local}}(x)$, we can make the first order gradient expansion:

$$f_{BE}(\omega_s, T^+) - f_{BE}(\omega_w, T^-) \approx \frac{\partial f_{BE}}{\partial T}(T^+ - T^-) = \frac{\partial f_{BE}}{\partial T}\Delta T_{\text{local}}(x)$$

The heat flux density (8) may thus be rewritten as:

$$Q = \frac{1}{2}\frac{\Omega}{(2\pi)^3}\sum_s \hbar\omega_s |v_s|\frac{\partial f_{BE}}{\partial T} \times \Delta T_{\text{local}}(x) = G_{\text{ballistic}}\,\Delta T_{\text{local}} \tag{12}$$

This formula (Eq. 12) is remarkable in the sense that for any homogeneous system, whatever the transport regime, the heat flux density is written as the product of the ballistic thermal conductance of the material and the local difference between the hemispherical temperatures. Hence, at the steady-state regime, the local temperature difference $\Delta T_{local}$ is uniform.

Besides, by combining **Erreur ! Source du renvoi introuvable.** with the effective conductivity $\kappa_{\text{effective}}$ defined by (6), it yields:

$$\Delta T_{\text{local}} = \Delta T_{\text{contacts}}\frac{\kappa_{\text{ballistic}}}{L\,G_{\text{ballistic}}} \tag{13}$$

Hence, in the case of a diffusive transport regime in which the phonon distribution remains close to equilibrium we have $\kappa_{\text{effective}} \ll L\,G_{\text{ballistic}}$ leading to $\Delta T_{\text{local}} = 0$ and $T^+ = T^- = T$, since at equilibrium all pseudo temperature definitions converge to the standard one. In the case of ballistic transport regime $\kappa_{\text{effective}} \gg L\,G_{\text{ballistic}}$ and thus $\Delta T_{\text{local}} = \Delta T_{\text{contacts}}$ i.e. the temperature gradients vanish: $T^+(x) = T^+(0)$ and $T^-(x) = T^-(L)$.

It must be emphasized that **Erreur ! Source du renvoi introuvable.** and its consequences are the key points of the presented work. Indeed, they are relevant not only in the case of a homogeneous system but also in the case of heterostructures as it is discussed in the next section. Importantly, they naturally solve the virtual interface paradox.

## III  Heterogeneous system

### III.1. Interfaces modeling

At each interface between two materials, a thermal boundary conductance $G^I$ appears expressed as follows:

$$Q = G^I \times \Delta T^I \tag{14}$$

where $\Delta T^I$ is the temperature drop at the interface. As discussed previously about the thermal conductivity, the definition of temperature is a crucial issue to use Eq. 14.

The common procedure uses the standard pseudo temperature, which is calculated from the full distribution of phonons, on each side of the interface. For an interface located at a position $x^I$, it gives:

$$\Delta T^I_{\text{ref}} = T(x^I - \delta x) - T(x^I + \delta x) \tag{15}$$

It should be however more consistent with the $G^I$ definition of [22] to use the hemispherical temperatures considering only the phonons that actually interact with the interface:

$$\Delta T^I_{\text{local}} = T^+(x^I - \delta x) - T^-(x^I + \delta x) \tag{16}$$

In the case of an out of equilibrium heat transfer, these two approaches are expected to significantly differ according to Eq. 12. To compare them, we have been used our home made phonon Monte Carlo simulator for phonons [14] that can manage all kinds of phonon distributions close to or far from equilibrium and can straightforwardly calculate the hemispherical temperatures. The studied heterostructures are homojunctions made of two identical Silicon films of various lengths separated by a rough (i.e. fully diffusive) interface. The transmission coefficient of each phonon colliding this diffusive interface is ½ according to the Diffusive Mismatch Model [27].

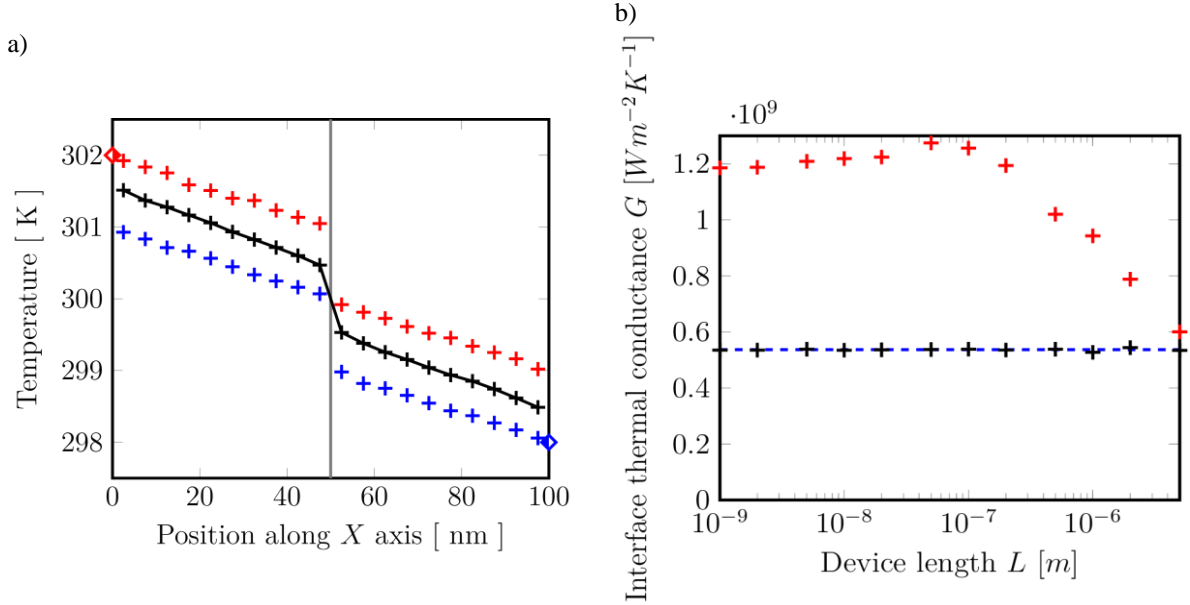

*Figure 2: In diffusive Si/Si junctions: a) temperature profiles T (black), T⁺ (red) and T⁻ (blue) for a length L = 100nm calculated by the Monte Carlo simulator, b) interface thermal conductance as a function of the device length. Analytical DMM result (blue dashed line), Monte Carlo results using Eq. 15 i.e. $\Delta T^I_{\text{ref}}$ (red crosses) and Eq. 16 i.e. $\Delta T^I_{\text{local}} = T^+ - T^-$ (black crosses)*

Figure 2.a) shows the temperature profiles of the pseudo temperature $T$ (black), and the hemispherical temperatures $T^+$ (red) and $T^-$ (blue) along the transport direction $x$ in a "heterojunction" of length $L = 100$ nm. The thermostats have temperatures of $T^{\text{hot}} = 302$ K and $T^{\text{cold}} = 298$ K, respectively. It is clear that the temperature differences $\Delta T^I_{\text{ref}}$ and $\Delta T^I_{\text{local}}$ are different and in this case $\Delta T^I_{\text{ref}}$ is almost two times smaller than $\Delta T^I_{\text{local}}$. This leads to different thermal interface conductances plotted in Figure 2.b). The thermal interface conductances obtained by using $\Delta T^I_{\text{ref}}$ (cf. Eq. 15) and $\Delta T^I_{\text{local}}$ (cf. Eq. 16) are plotted in red and black crosses, respectively. For comparison, the results provided by the standard DMM formula are indicated by a blue dashed line. The results obtained using temperature difference of $\Delta T^I_{\text{ref}}$ exhibits a strong and unexpected length-dependence. As expected in long devices where the transport is close to equilibrium, the two approaches tend to become similar. Remarkably, the results obtained from the use of the $\Delta T^I_{\text{local}}$ temperature difference reproduce the analytical DMM results for all device lengths.

One of the important advantages of the formulation with $\Delta T^I_{\text{local}}$ for considering the interface conductance is that the virtual interface paradox disappears. Indeed, considering a transparent (i.e. ballistic) interface located anywhere in the material, $\Delta T^I_{\text{local}} = \Delta T_{\text{local}} = T^+ - T^-$ and the corresponding interface thermal conductance $G^I = G_{\text{ballistic}}$ is finite. For all these reasons, only this method will be used in the following.

### III.2. Analytical model of thermal transport in heterojunctions

A complete analytical model of thermal transport in a heterojunction can now be derived. Typical profiles of temperatures $T$, $T^+$ and $T^-$ in a heterostructure are plotted in Figure 3 where all the relevant parameters are also indicated. A superscript $L$ stands for the left part of the junction and $R$ for the right part. The total temperature difference between the left thermostat at a temperature $T^{\text{hot}}$ and the right thermostat at a temperature $T^{\text{cold}}$ can be decomposed as follows:

$$T^{\text{hot}} - T^{\text{cold}} = [\Delta T^L_{\text{contact}} - \Delta T^L_{\text{local}}] + \Delta T^I + [\Delta T^R_{\text{contact}} - \Delta T^R_{\text{local}}] \quad (17)$$

Besides, as the thermal flux is uniform and using our model described above (cf. Eq. 6, 14 and 5, respectively), we can write:

$$Q = \frac{\kappa^{L/R}_{\text{effective}}}{L^{L/R}} \Delta T^{L/R}_{\text{contact}} = G^I \cdot \Delta T^I = G^{L/R}_{\text{ballistic}} \cdot \Delta T^{L/R}_{\text{local}} \quad (18)$$

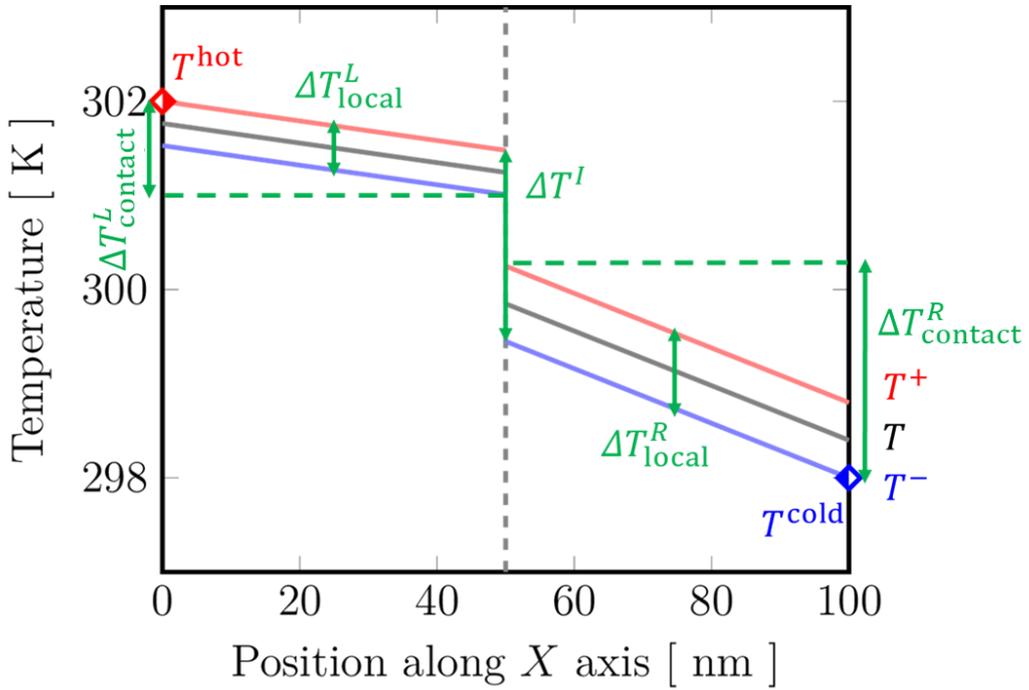

*Figure 3: Schema of temperature profiles: T (black), $T^+$ (red) and $T^-$ (blue) in a heterojunction. Diamonds show the temperature of hot (red) and cold (blue) thermostats.*

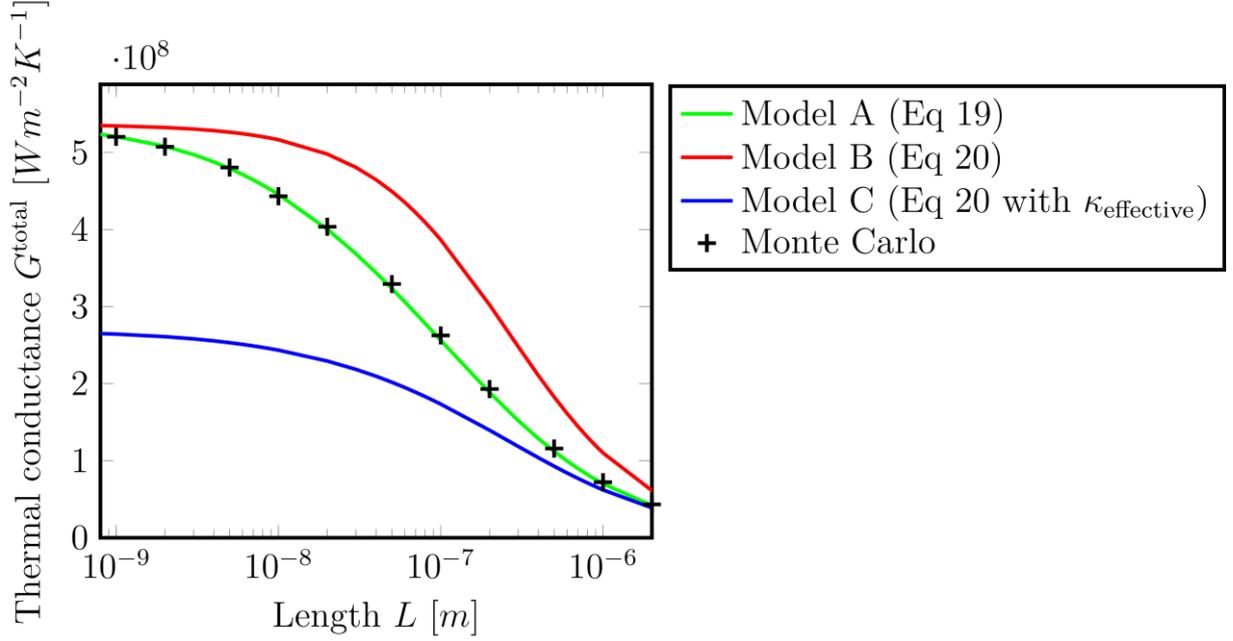

*Figure 4: Total thermal conductance of diffusive Si/Si junctions of different length from MC simulations (crosses) and several analytical models (lines).*

Finally, the total conductance of the heterojunction $G^{\text{total}}$ is expressed as a function of the interface conductance $G^I$, the effective conductivities of involved materials and also their ballistic conductances, i.e.

$$G^{\text{total}} = \frac{Q}{T^{\text{hot}} - T^{\text{cold}}} = \left[ \frac{L^L}{\kappa^L_{\text{effective}}} - \frac{1}{G^L_{\text{ballistic}}} + \frac{1}{G^I} - \frac{1}{G^R_{\text{ballistic}}} + \frac{L^R}{\kappa^R_{\text{effective}}} \right]^{-1} \quad (19)$$

As a first approach of conductance estimation, the use of the conventional non-spectral Matthiessen approximation (cf. Eq (8)) for the effective conductivities yields:

$$G^{\text{total}}_{\text{standard}} = \left[ \frac{L^L}{\kappa^L_{\text{diffusive}}} + \frac{1}{G^I} + \frac{L^R}{\kappa^R_{\text{diffusive}}} \right]^{-1} \quad (20)$$

Interestingly, the common expression of the total conductance obtained when considering the standard model of three thermal resistances in series, including the two resistances of the left and right films (defined by using the standard bulk diffusive thermal conductivity of materials) plus the interface resistance, is recovered. However as shown in Figure *1*, this non-spectral Matthiessen approximation is disappointing in the intermediate transport regime requiring better approximation of the effective thermal conductivity.

In the case of the diffuse Si/Si junctions previously presented in section 1.6, the total conductances provided by the previous models are compared in Figure 4 with their numerical counterpart provided by our Monte Code. Three analytical models were investigated. The first model called A, indicated in green solid line, is the more complete one is based on Eq. 19 including the semi analytical effective conductivities of Eq. 9. The second one called B, in red, corresponds to the classic 3 thermal resistances in series approach of Eq 20. The third model called C is also based on Eq. 20 but the semi-analytical effective conductivities $\kappa^{L/R}_{\text{effective}}$ of Eq. 9 replace the bulk diffusive conductivity $\kappa^{L/R}_{\text{diffusive}}$.

Counterintuitively, the model C is very disappointing as it strongly underestimates the conductance and cannot capture the ballistic regime. The use of effective conductivities are not a sufficient ingredient to investigate efficiently the out of equilibrium regime. Model B is more satisfactory but it overestimates by over 50% the MC results in the intermediate regime. Model A, our new model, is in agreement with the MC simulation with errors lower than 5% over the entire length range.

# IV Conclusion

A comprehensive model able to accurately describe the heat transfer through heterostructures in all phonon transport regimes has been presented.

First, in homogenous materials, the concepts of ballistic conductance, useful in the context of fully ballistic systems, and of effective conductivity relevant in both ballistic and diffusive regime as well in the intermediate one, have been presented. It should be noted that these parameters are related to non-local temperature difference and are geometry dependent while the standard Fourier thermal conductivity is defined locally by using temperature gradient and is not geometry dependent.

Second, the interface thermal conductance has been investigated by introducing two hemispherical temperatures $T^+$ and $T^-$ distinguishing the phonon populations according to the direction of their velocity. It appears to be an appealing solution to make the interface thermal conductance modeling consistent with Monte Carlo simulations. Combined with the previous concepts, the historic virtual interface paradox is solved.

Finally, a comprehensive analytical framework based on the central relationship: $Q = G_{\text{ballistic}} (T^+(x) - T^-(x))$ has been derived. This framework extends the validity range of previous models and can reproduce accurately results of numerical Monte Carlo simulation in Si homojunctions in all transport regimes: ballistic, diffusive, and also intermediate ones. This versatile and easy to use approach should be particularly suitable to investigate heat transfer in various kinds of nanostructures.